\documentstyle[aps,eqsecnum,epsf,floats]{revtex} 
\begin{document}
\draft
\twocolumn[\hsize\textwidth\columnwidth\hsize\csname
           @twocolumnfalse\endcsname
\def\epsfxy5020{\epsfxsize=5.0in \epsfysize=2.0in}

 \def\picboxnormal{(3.75,2.0)}
\def\epsfxy{\epsfxsize=4.5in \epsfysize=3.18in}

\title{ Dynamical Evolution of Boson Stars in Brans-Dicke Theory}
\author{
Jayashree Balakrishna\cite{bala} and Hisa-aki Shinkai\cite{his} }
\address{Department of Physics, Washington University, 
St. Louis, MO 63130-4899, USA }

\date{February 6, 1998 (revised version)}
\maketitle
\begin{abstract}
\widetext
We study the dynamics of a self-gravitating scalar field 
solitonic object (boson star) in
the Jordan-Brans-Dicke (BD) theory of gravity. 
We show dynamical processes of this system such as 
(i) black hole formation of perturbed equilibrium configuration on
an unstable branch; 
(ii) migration of perturbed equilibrium configuration 
from the unstable branch to stable branch;  
(iii) transition  from excited state to a ground state.  
We find that the dynamical behavior of boson stars in BD theory 
is quite similar to that in general relativity (GR),
with comparable scalar wave emission.
We also demonstrate the formation of a stable boson star 
from a Gaussian scalar field packet 
with flat gravitational scalar field initial data. 
This suggests that boson stars can be formed 
in the BD theory in much the same way as in GR. 
\end{abstract}
\pacs{PACS number(s): 04.40.-b, 04.40.Dg, 04.50.+h~~~~~~~~~
to appear in {\it Phys. Rev. D}}

\vskip 2pc]
\narrowtext


\section{Introduction} \label{sec1}
Gravitational solitonic objects are quite an interesting topic in general
relativity (GR).  
A boson star consists of massive complex scalar field, and
was first discussed by Kaup \cite{K} 
and then by Ruffini and Bonazzola \cite{RB} (for a thorough review, 
see \cite{LM,BSSreview}).  They can form stable configurations 
having negative binding energy, as a result of a balance of the dispersion
due to the classical analog of the uncertainity principle and
the attractive effects of gravity. 
If we include even a small self-interaction term, then 
their maximum allowed stable mass can be close to the order of a solar mass 
\cite{CSW}.  It is also speculated that they are a  form of dark matter
that could have been created during a phase transition in the early 
universe (see Frieman et al \cite{FGGK}).  
Although we have still no evidence for their astrophysical existence, 
these systems are a good model to learn the nature of a strong 
gravitational field.

The stability of boson star has also been studied by several authors.  
Lee and Pang \cite{leepang} discussed ground state stability using 
linearized perturbation theory, and Seidel and Suen \cite{SS90} studied 
their dynamical behavior by evolving field equations numerically.  
In GR, 
the ground state boson star configurations  comprise of a stable branch
and an unstable branch.
Upon perturbations, boson stars on the stable branch remain in the
stable branch, settling down into a configuration with a different
mass.  In the process, it emits scalar radiation with some
characteristic normal mode frequencies. 
On the other hand, stars on the unstable branch do not remain
these after perturbations. They either disperse
completely, form black holes or migrate to the stable branch 
depending on the size of the perturbations. 
These qualitative features are also discussed applying catastrophe
theory\cite{cata}. 
Recently, Balakrishna, Seidel and Suen \cite{BSS} studied
dynamical boson stars with the self-coupling term and excited states.
They found that excited state boson star equilibrium configurations
have branches similar to that of grand state, but all branches are
unstable.

In this paper, 
we study the dynamical behaviour of boson stars
in the Jordan-Brans-Dicke (BD) theory of gravity\cite{BD}.
In particular, we compare the dynamics 
with those in GR.
The BD theory is one of the alternative theories of gravity to GR, 
and the most simple and proto-type in all the scalar-tensor 
theories of gravity (see Will \cite{CW} for a review). 
The previous experimental test using the delay of radar echo in the solar
system shows the bound of the BD parameter as $\omega_{BD} > 500$
\cite{VIKING,VLBI}, of which the infinite limit agrees with GR.  
This bound is also considered to be limited by the direct observations of 
gravitational waves (see \cite{SSM} and the references there in). 
[Recently, more strict limitations of the  BD parameter $\omega_{BD}$
are reported \cite{nucleo}, however the results are model
dependent and we think $\omega_{BD} > 500$ is still the generally 
accepted observational limit.]

So far, boson stars in scalar-tensor gravity have also been discussed by
Gundersen and Jensen \cite{GJ} and  Torres \cite{torres}
showed the existence of equilibrium ground state boson star 
solutions in the BD coupling, and in the three different couplings 
in the scalar-tensor theories, respectively. 
Comer and Shinkai \cite{CS97} showed the existence of excited state
boson stars in both BD and Damour-Nordtvedt's quadratic coupling 
(attractor) model \cite{DN} in the scalar-tensor theories. 
They also discussed the stabilities of ground state boson 
stars using catastrophe theory.
One of our purposes in this paper is to study the stability of boson 
stars in BD theory. 
By evolving the field equations numerically for slightly or heavily perturbed
equilibrium data, we clarify the stability of boson star both for
ground and excited states. 

Our system includes two scalar fields:  
bosonic matter (complex and massive) scalar field and gravitational
(real and massless) scalar field  (which hereafter we call BD field). 
We expect to be able to study the fundamental mechanism of
the interactions between these two fields
in their dynamics. 
Several previous simulations have shown emissions of scalar waves in BD 
\cite{matsuda,SNN,SST} or scalar-tensor theory \cite{harada,novak}
from a collapse of dust or star model. 
Our second interest is in how
much difference appears in the scalar gravitational wave
emissions during the dynamical boson star system  between BD and GR:
whether they will be enhanced or supressed.

Throughout this paper, we stand at the point 
to see if they in any way differ from boson stars in a detectible
manner; that is, we impose $\omega_{BD}>500$ in most simulations. 

The outline of this paper is as follows. 
In Sec. \ref{sec2}, we will introduce the field equations, and 
basic outline of our numerical techniques.  With a view to study
this problem in scalar-tensor theories
in the near future \cite{BCS97b},
we describe fundamental equations not only in BD theory but in general
scalar-tensor theories. 

We describe the equilibrium configuration briefly in Sec. 
\ref{sec:equili}.  We plot the sequences of excited states 
configurations in BD theory and also discuss fractional anisotropy
measurements in this system, both of which have not appeared in the 
references. 

Sec. \ref{sec:stability} is devoted for stable boson star configurations.
In GR, under the effects of finite perturbations, these stars react by 
expanding and contracting with emitting scalar radiation at each expansion 
\cite{SS90}. The star looses mass and settles to a lower mass 
configuration.  Each expansion (contraction)
of the core of the star is accompanied by the contraction (expansion)
of the radial metric. 
We study whether the expansions and 
contractions of the star set
the BD field into oscillations and whether this results in any measurable
gravitational radiation in addition to the scalar radiation.
Under infinitesimal perturbations, boson stars in GR start 
oscillating at their fundamental quasinormal mode frequencies that remain
constant and virtually undamped for large periods of time. 
We also study whether this is seen in BD theory. 

Sec. \ref{sec:transition} shows the behavior of unstable boson stars.
In GR, equilibrium boson stars on the unstable branch, when perturbed, 
begin forming black holes or migrating to a new equilibrium
configuration on the stable branch.
The excited state configurations of boson stars are expected to be
unstable. If they cannot
lose enough mass and settle to the ground state configuration they 
collapse to black holes.  We also show the cascading of a BD boson star in
an excited state to the ground state.

In Sec. \ref{sec:formation}, 
we look at the formation of boson stars in  BD theory.
The collection of bosonic matter localized in a region of space is
represented by a Gaussian initial boson field.  
The BD field
itself is initially set to zero,  
we  discuss whether or how an equilibrium 
configuration actually forms in BD theory.
Finally, in Sec. \ref{sec:conclude} we make some 
concluding remarks.  

We use the units $c = 1$ and $\hbar = 1$. 
This implies that the scalar field mass $m$ is an inverse length 
(actually, the inverse Compton wavelength of the scalar particles) and 
the bare gravitational constant $G_{*}$ has units of length squared.  

\section{The field equations}\label{sec2}

In this section, we present our basic field equations,
 boundary conditions,
and numerical techniques to solve the system.  
For our future convenience \cite{BCS97b}, 
we show the field equations
not only in the Brans-Dicke theory of gravity 
($\omega=\omega_{BD}=const.$), 
but also in the general 
scalar-tensor theory of gravity ($\omega=\omega(\phi)$). 

\subsection{Lagrangians and field equations}\label{sec2-1}
The action for our system of scalar-tensor gravity coupled to a 
self-interacting, complex scalar field in the physical, ``Jordan-frame'' 
is
\begin{eqnarray}
    S_J &=& {1\over 16 \pi} \int d^4x  \sqrt{-\tilde{g}} 
          \left[\phi \tilde{R} -\phi^{-1} \omega(\phi) 
          \tilde{g}^{\mu\nu}  \partial_\mu \phi \partial_\nu \phi
          \right] \nonumber \\
        && - 
         \int {\rm d}^4x \sqrt{- \tilde{g}} 
          \left[{1 \over 2} \tilde{g}^{\mu \nu} \partial_{\mu} 
          \psi^{\dag} \partial_{\nu} \psi +  V\left(\psi^{\dag} \psi\right)
          \right]. \label{2.1} 
\end{eqnarray}
The gravitational scalar is $\phi$ and $\omega(\phi)$ is the Jordan-frame
coupling of $\phi$ to the matter.  The complex scalar $\psi$ (with its 
complex conjugate being $\psi^{\dagger}$) has mass $m$ and is 
self-interacting through the potential 
\begin{equation}
V\left(\psi^{\dag} \psi\right)={m^2 \over 2} 
          \psi^{\dag} \psi +{\Lambda \over 4} 
         ( \psi^{\dag} \psi )^2. \label{potent}
\end{equation}
The strength of the self-interaction, $\Lambda$, is normally taken to be 
positive. 
 
There is an alternative representation of the action above, the   
so-called ``Einstein-frame.''  
The expression is given by the conformal transformation 
\begin{equation}
   \tilde{g}_{\mu \nu} =  e^{2 a(\varphi)} g_{\mu \nu} \ , \label{2.2}
\end{equation}
where $a(\varphi)$ is the functional transformation from $\phi$ to the 
Einstein-frame gravitational scalar $\varphi$, 
\begin{equation}
    \phi^{- 1} = G_* e^{2 a(\varphi)}, \label{2.3}
\end{equation}
where  $G_*$ is the effective gravitational constant in the 
Einstein-frame.
The relationship 
between $\omega(\phi)$ and $a(\varphi)$ is obtained from
\begin{equation}
    \alpha^2 = (2 \omega + 3)^{- 1} \ , \label{2.4}
\end{equation}
where
\begin{equation}
   \alpha(\varphi) \equiv {\partial a \over \partial \varphi} \ . 
        \label{2.5}
\end{equation}
The action in the Einstein-frame is thus
\begin{eqnarray}
&&  S_E = {1 \over 16 \pi G_{*}} \int {\rm d}^4x \sqrt{-g} \left[R - 
      2 g^{\mu \nu} \partial_{\mu} \varphi \partial_{\nu} \varphi 
      \right] \nonumber \\ &&
     - \int {\rm d}^4x \sqrt{-g} \left[{1 \over 2} 
      e^{2 a(\varphi)} g^{\mu \nu} \partial_{\mu} \psi^{\dag} 
      \partial_{\nu} \psi + e^{4 a(\varphi)} 
      V\left(\psi^{\dag} \psi\right) \right] 
      \ . \label{2.6}
\end{eqnarray}
It does not deliver GR exactly because the 
metric $g_{\mu \nu}$ is not the true, physical metric that encodes the 
distance between spacetime points.  However, the Einstein-frame 
does deliver equations that are similar enough to GR 
that we will use it for our calculations. 

The Einstein-frame stress-energy tensor is
\begin{eqnarray}
   T_{\mu \nu} &=& {1 \over 2} e^{2 a(\varphi)} \left(\partial_{\mu} 
                \psi^{\dag} \partial_{\nu} \psi + \partial_{\nu} 
                \psi^{\dag} \partial_{\mu} \psi\right)\nonumber \\
               && - {1 \over 2} 
                e^{2 a(\varphi)} \left( \partial_{\tau} \psi^{\dag} 
                \partial^{\tau} \psi + 2 e^{2 a(\varphi)} 
                 V(\psi^{\dag} \psi)\right) 
                g_{\mu \nu} \ .
\end{eqnarray}
The gravitational field equations for $g_{\mu \nu}$ and $\varphi$ are
\begin{equation}
   G_{\mu \nu} = 8 \pi G_{*} T_{\mu \nu} + 2 \partial_{\mu} \varphi 
               \partial_{\nu} \varphi -  \partial_{\tau} \varphi 
               \partial^{\tau} \varphi  g_{\mu \nu}
\label{einsteineq}
\end{equation}
and
\begin{equation}
   \nabla_{\sigma} \nabla^{\sigma} \varphi = - 4 \pi \alpha T \ , 
\end{equation}
where $T$ is the trace of the stress-energy tensor.  The matter field 
equations are
\begin{eqnarray}
   \nabla_{\sigma} \nabla^{\sigma} \psi^{\dag} + 2 \alpha 
                \partial_{\tau} \psi^{\dag} \partial^{\tau} \varphi 
               & =& 2 e^{2 a(\varphi)} {\partial 
                V \over \partial \psi}, \label{2.10a}\\
   \nabla_{\sigma} \nabla^{\sigma} \psi + 2 \alpha  \partial_{\tau} 
                \psi \partial^{\tau} \varphi &=& 2 e^{2 a(\varphi)} 
                {\partial V \over \partial 
                \psi^{\dag}}\ . \label{2.10b}
\end{eqnarray}

The coupling function $a(\varphi)$ is given by choosing a
theory of gravity.  
In this paper, we only consider the Brans-Dicke coupling
\begin{equation}
    a(\varphi) = {\varphi - \varphi_{\infty} \over \sqrt{2 \omega_{BD} 
                 + 3}}, \label{2.11}
\end{equation} 
where the parameter $\omega=\omega_{BD}$ is constant, which
observational constraint is known as $\omega_{BD} > 500$ 
\cite{VIKING,VLBI}. 
The term 
$\varphi_{\infty}$ represents the asymptotic value of the gravitational 
scalar field.  

 Because the potential $V\left(\psi^{\dag} \psi\right)$ is 
a functional of $\psi^{\dag} \psi$,   it preserves the global U(1) 
gauge symmetry ($\psi \to e^{i \sigma} \psi$, where $\sigma$ is a 
constant) present in the theory.  This symmetry results in a conserved 
current, whose explicit form in the Jordan-frame is
\begin{equation}
    \tilde{J}^{\mu} = {i \over 2} e^{-2 a(\varphi)} g^{\mu \nu} 
              \left(\psi \partial_{\nu} \psi^{\dag} - \psi^{\dag} 
              \partial_{\nu} \psi\right) \ . \label{2.14}
\end{equation}
This conserved current leads to a conserved charge, which is $N_p$, 
the number of particles making up the star:
\begin{equation}
  N_p=\int d^3x \sqrt{-g}  \tilde{J}^{t}.
\end{equation}

The spacetime considered here is spherically symmetric, 
with the Einstein-frame metric taking the form
\begin{equation}
     {\rm d}s^2 = - N^2(t,r) {\rm d}t^2 + g^2(t,r){\rm d}r^2 + r^2 
                  \left[{\rm d}\theta^2 + {\rm sin}^2 \theta {\rm d}
                  \phi^2\right],  \label{3a.1}
\end{equation}
where $N$ is the lapse function and $r$ is the circumferential
radius.  We dropped the shift vector, since we use a polar-slicing 
condition \cite{bard}  in evolution. 

In this coordinate system, 
the Jordan-frame ADM mass $M_J$ is 
given by
\begin{equation}
    G_{*} M_J = \lim_{r \to \infty} {r \over 2} \left(1 - 1/
              \tilde{g}_{r r}\right) \ . \label{3c.2}
\end{equation}
The similar Einstein-frame ADM mass $M_E$ is
\begin{equation}
    G_{*} M_E = \lim_{r \to \infty} {r \over 2} \left(1 - 1/g_{r r}
              \right) \ . \label{3c.3}
\end{equation}
However, since $\tilde{g}_{r r} = e^{2 a(\varphi)} g_{r r}$ and we set 
$a(\varphi_{\infty}) = 0$ [to be discussed below], 
then the limits on the right-hand-sides are 
equal and therefore $M_J = M_E \equiv M$.

\subsection{The equilibrium state equations} \label{sec2-equili}

The gravitational scalar, which is real, is assumed also to be 
spherically symmetric and static when we solve the equilibrium 
configuration:
\begin{equation}
     \varphi = \varphi(r) \ . \label{3a.2}
\end{equation}
As for the matter scalar field, Friedberg et al \cite{FLP} show that 
the minimum energy configurations are those for which 
\begin{equation}
     \psi = e^{- i \Omega t} \Phi(r) \ , \label{3a.3}
\end{equation}
where $\Omega$ is real and positive and $\Phi(r)$ is real function. 
 Their proof (see the Appendix 
in \cite{FLP}) also goes through for scalar-tensor gravity, and so we 
will take $\psi$ to have this form. 
 
We will take advantage of scale-invariances of the field equations to 
redefine some of the fields, parameters, and the radial and time 
coordinates:
\begin{eqnarray}
&& mr \to r, 
\sqrt{4 \pi G_{*}} \Phi \to \Phi,  m N/\Omega \to N, \nonumber \\&&
\Lambda/4 \pi G_{*} m^2 \to \Lambda, \Omega t/m \to t. \label{3e.1}
\end{eqnarray}
Note that the rescaling changes the asymptotic value of $N$, 
which is now 
\begin{equation}
    \lim_{r \to \infty} N(x) = m/\Omega \ . \label{3e.4}
\end{equation}

The field equations, then, become 
\begin{eqnarray}
{\partial_r}{\partial_r}\varphi&=& 
[-{g^2 +1 \over r} + 2 e^{4a} r g^2 V(\Phi)]
{\partial_r}\varphi \nonumber \\&&
+ g^2 \{ 2 \alpha e^{2a}
 [{1 \over 2} \nabla_\sigma \Phi  \nabla^\sigma \Phi
+2e^{2a} V(\Phi) ]\}, \label{field1-slv} \\
{\partial_r}{\partial_r} \Phi&=& 
[-{g^2 +1 \over r}  +  2 e^{4a} r g^2 V(\Phi)] 
{\partial_r} \Phi
 -{g^2 \over N^2} \Phi  \nonumber \\&&
+ 2 g^2 [e^{2a} {d V(\Phi) \over d \Phi} - \alpha
\nabla_\sigma \Phi  \nabla^\sigma \varphi], \label{field2-slv} \\
{\partial_r} (g^2)&=& {g^2} ({g^2 r \over N^2} T_{00}
- {g^2 -1 \over r} ),  \label{field3-slv} \\
{\partial_r} (N^2)&=& {N^2} (r T_{11} + {g^2 -1 \over r} ),
\label{field4-slv} 
\end{eqnarray}
where 
\begin{eqnarray}
T_{00}&=&{N^2\over g^2} ({\partial_r}\varphi)^2  \nonumber \\&&
+e^{2a} [ \Phi^2 +{N^2 \over g^2} ({\partial_r} \Phi)^2
 +2 N^2 e^{2a}
V(\Phi)], \\
T_{11}&=& ({\partial_r}\varphi)^2  \nonumber \\&&
+e^{2a} [ ( {\partial_r} \Phi )^2
+{g^2 \over N^2} \Phi^2 -2 g^2 e^{2a}
V(\Phi)].
\end{eqnarray}


The boundary conditions for this system of equations must take into 
account three things: (i) the solutions must be geometrically 
regular at the origin; (ii) the solutions must yield an asymptotically 
flat spacetime; and (iii) the solutions must take into account the 
cosmological input for both the coupling $a(\varphi)$ as well as 
$\varphi$.  

Geometrical regularity at the origin means there is no conical 
singularity, i.e., the proper radius divided by the proper circumference 
should reduce to $2 \pi$ at $r = 0$.  This implies that $g(0) = 1$.  
Also, to maintain regularity in the field equations as $r \to 0$, we 
impose that $d\Phi / dr |_{r=0} = 0$ and $d \varphi / dr |_{r=0} = 0$.    

For a purely technical reason to set $M_J = M_E$ , we desire 
solutions that are asymptotically flat in both the Jordan and 
Einstein frames.  That is, we want both $\tilde{g}_{\mu \nu}$ and 
$g_{\mu \nu}$ to reduce to the flat spacetime metric at spatial 
infinity.  The implication of this is that the value of 
$\varphi_{\infty} \equiv \varphi(\infty)$ must be such that 
$a(\varphi_{\infty}) = 0$. 
This is guaranteed since 
there is one more rescaling that has no analog in 
GR. 
That is an invariance of the field equations 
if an arbitrary constant is added to the scalar-tensor coupling.  If we 
simultaneously do the rescaling
\begin{equation}
   e^c x \to x \ , \ e^c \Phi \to \Phi \ , \ e^c N \to N \ , \
         e^c \Lambda \to \Lambda \  \label{3e.6}
\end{equation}
on the variables defined by Eq. (\ref{3e.1}) and let $a(\varphi) + c 
\to a(\varphi)$, then the field equations remain unchanged.  

For the BD coupling, $\Phi_c \equiv \Phi(0)$ and 
$\varphi_{\infty}$ are the only freely specified field values.  The 
value of $N(0)$ is not specified freely, but rather is determined so 
that $\Phi(\infty) = 0$.  The value of $\varphi$ at the origin is not 
specified freely; it must be determined in such a way that the solution 
for $\varphi$ goes to $\varphi_{\infty}$ at spatial infinity.  
We will 
use the freedom to add an arbitrary constant to the BD coupling 
$a(\varphi)$ so that all the 
solutions we consider have $\varphi_{\infty} = 0$.

\subsection{The evolution equations} \label{shinkai-note}

We, here, assume the gravitational scalar field is time dependent
$\varphi=\varphi(t,r)$ and use the rescaled bosonic field $\Psi$ as
$\Psi=\sqrt{4\pi G_*}\psi$. 
Analogous to eq.(3.6)-(3.10) in \cite{SS90},
we introduce  scalar field momenta $\Pi_\varphi$ and $\Pi_\Psi$:
\begin{eqnarray}
\Pi_\varphi&=& {g \over N} \partial_t (r \varphi)
\equiv {1\over \beta} \partial_t (r \varphi), \\
\Pi_\Psi&=& {g \over N} \partial_t (r\Psi) \equiv {1\over \beta}
 \partial_t (r\Psi),
\end{eqnarray}
where we set $\beta=N/g$.
The field equations  become 
\begin{eqnarray}
\partial_t (r \varphi) &=&  
\beta \Pi_\varphi,  \label{dyn-1} \\
\partial_t \Pi_\varphi 
&=& (\partial_r\beta) \partial_r(r\varphi) 
+\beta \partial_r\partial_r(r\varphi)
 -(\partial_r\beta)
(r\varphi){1\over r} \nonumber \\&&
 - Ngr~
2\alpha e^{2a}
 [{1 \over 2} \nabla_\sigma \Psi  \nabla^\sigma \Psi^\dagger
+2e^{2a} V(\Psi\Psi^\dagger) ], \label{dyn-2}  \\
\partial_t (r\Psi) &=& 
\beta \Pi_\Psi,  \label{dyn-3} \\
\partial_t \Pi_\Psi 
&=& (\partial_r\beta) \partial_r(r\Psi) + \beta \partial_r\partial_r(r\Psi)
 -(\partial_r\beta)(r\Psi){1\over r}  \nonumber \\&&
- 2 Ngr [ e^{2a} {d V(\Psi\Psi^\dagger) \over d \Psi^\dagger} - \alpha
\nabla_\sigma \Psi  \nabla^\sigma \varphi] \label{dyn-4}.
\end{eqnarray}
Note that  $\Psi$ and $\Pi_\Psi$ are complex variables, so
(\ref{dyn-3}) and (\ref{dyn-4}) have two components.
The momentum constraint and the $G_{rr}$ component of the Einstein 
equations become
\begin{eqnarray}
\partial_t g &=&
N [ \Pi_\varphi \partial_r\varphi +  e^{2a} {1\over 2}
(\Pi_\Psi^\dagger \partial_r\Psi+\Pi_\Psi \partial_r\Psi^\dagger)], 
\label{momco} \\
\partial_rN &=& {N\over 2r}(g^2-1) + {Nr \over 2} \left(
(\partial_r\varphi)^2 + \Pi_\varphi^2 {1\over r^2} 
\right.  \nonumber \\&& \left.
+ e^{2a} [(\partial_r\Psi)(\partial_r\Psi^\dagger) 
+\Pi_\Psi\Pi_\Psi^\dagger {1\over r^2} -2 g^2 e^{2a}
V(\Psi\Psi^\dagger)]
 \right). \label{branlap}
\end{eqnarray}

We use the above set of equations (\ref{dyn-1})-(\ref{branlap}) 
for evolving the system and use the Hamiltonian constraint equation
\begin{eqnarray}
&& 2{\partial_r g \over gr} +{g^2-1 \over r^2}=
{\Pi_\varphi^2 \over r^2} +(\partial_r\varphi)^2 \nonumber \\&&
+e^{2a}[ {\Pi_\Psi\Pi_\Psi^\dagger \over r^2}
 +(\partial_r\Psi)(\partial_r\Psi^\dagger)
 +2 g^2 e^{2a} V(\Psi\Psi^\dagger) ]
\label{branham}
\end{eqnarray}
to check the accuracy of our simulation. 

\subsection{Numerical techniques} \label{sec3f}
\subsubsection{Equilibrium configurations} 

We use fourth-order Runge-Kutta algorithm to solve 
differential equations (\ref{field1-slv})-(\ref{field4-slv}).  
In order to find an equilibrium configuration, 
our system requires a two parameter search to find a solution
that satisfies the boundary conditions for both $\Phi_\infty$ and
$\varphi_\infty$.  Operationally, we choose a central value of the
scalar field $\Phi_c$ first together with a guessed central value of the
gravitational scalar field $\varphi(0)$, and integrate out to large radii
for different values of $N(0)$.  We then check if the resulting
$\varphi_{\infty}$ is close to our expected boundary value.

The falling off behaviour
for the gravitational scalar field $\varphi$ is much slower than the matter
scalar field $\Phi$. Therefore, at the numerical boundary, say
$r=r_{end}$, we set the expected boundary value for $\varphi(r_{end})$
as  $\varphi(r_{end})=\varphi_{\infty}+{C / r_{end}}$, where a constant
$C$ is given by $C=-r_{end}^2 \times \left. {d \varphi(r) \over d r}
\right|_{r=r_{end}}$.
If the computed $\varphi(r_{end})$ is not the expected value,  then we
change $\varphi(0)$ and repeat the whole procedure.  We set the tolerances
to judge convergence in $\varphi(r_{end})$ as $5 \times 10^{-7}$.
More details are in \cite{CS97}.

\subsubsection{Evolutions}
We use the same code that was used in \cite{SS90} for
evolutions  with modifications to incorporate  BD theory. A polar slicing
condition \cite{bard} for the lapse is hard wired into the code. 
This slicing is highly singularity
avoiding and in the event of formation of an apparent horizon, the lapse
rapidly collapses and the radial metric blows up crashing the code
indicating imminent black hole formation.
The lapse equation (\ref{branlap}) is integrated once on every time 
slice using a sixth
order integration scheme. The Hamiltonian constraint equation
(\ref{branham}) is monitored as an indicator of the accuracy of the
simulation and is not solved during the evolution. 
A leapfrog evolution scheme is used as described in
\cite{BSS}. 

For the boundary conditions regularity dictates that the
radial metric is equal to 1 at the origin. The boson field and the 
BD field are both specified at the origin. The boson field goes
to zero at $\infty$ and the BD field goes to a constant which is
fixed during the evolution. 
This constant does not enter into any
of the evolution equations as all the terms in the set of equations
are derivatives in the BD field.

The inner boundary at the origin requires that 
the derivatives of all the metrics and fields vanish
at this point.  This is implemented
by extending the range of $r$ to include negative values. 
The metric compnents $g$, $N$,
the boson field and the BD field are required to be symmetric
about $r=0$. 
The code itself uses new field variables which are the original 
fields times $r$ for both the boson fields as 
well as the BD field 
and the momentum variables $\Pi$ described in the
previous section. Thus the code field
variables as well as $\Pi_{\varphi}$ and $\Pi_{\Psi}$ are antisymmetric
about $r=0$. The bosonic and BD fields at the origin during the 
evolution are determined from the first derivatives of the code fields
there.
For the radial metric the value at the boundary is determined from the
Hamiltonian constraint equation (\ref{branham}).

In order to prevent reflections from the edge of the grid we use an 
{\it outgoing wave} condition at the outer edge of the grid 
(far away from the star) for the matter field $\Psi$ and 
the BD field $\varphi$. 

For the BD field, $\varphi$, 
we impose the wave behaves as 
\begin{equation}
\varphi(r,t) = \varphi_\infty + {1\over r}F(t-{r\over c}) \label{e:ccfi}
\end{equation}
which is a natural out-going wave condition in spherically 
flat spacetime, and which is the proper assumption for the 
asymptotic region.  
Differentiating, one gets
\begin{equation}
\left .
{1\over c}{\partial \varphi \over \partial t} + {\partial \varphi \over
\partial r} + {(\varphi-\varphi_\infty) \over r} \, 
\right |_{\text{outer edge}} = 0.
\end{equation}
Note that this is the same technique used by Novak \cite{novak}, who
studied a stellar collapse in scalar-tensor theory. 

For the boson field, $\Psi$,  an asymptotic solution of the form 
$e^{-k\,r}e^{i\,\Omega\,t}/r$
to order $\frac{1}{r}$ is assumed. 
This gives an outgoing boundary condition to this order of
\begin{equation}
\beta^2 k^2=\Omega^2-N^2m^2 e^{2a}.
\end{equation}
This dispersion relation is nontrivial for a massive scalar field. There
is no perfect algorithm to implement it. At the outermost grid
point we require that
\begin{equation}
\partial_t \partial_t \tilde{\Psi} = 
-\beta \partial_t \partial_r \tilde{\Psi} -\frac{N^2}{2}
\tilde{\Psi} e^{2a},
\end{equation} 
where $\tilde{\Psi} =r\,\Psi$.  
The second term on the right is a finite mass correction to leading
order.

In addition to remove second order reflections we put a sponge 
for the matter field, (it reduces the momentum of the boson field
artificially, and is irrelevant for the massless BD field) 
which is a potential term that is large for incoming waves
 (proportional to $k+\Omega$) and small for outgoing waves
(proportional to $k-\Omega$).  
Therefore, 
we add an additional term in the evolution equation for $\Pi_\Psi$
(\ref{dyn-4})
\begin{equation}
 \frac{V(\Psi\Psi^\dagger)}{e^{a}N}(\Pi_\Psi+\partial_r \tilde{\Psi})
\end{equation}
for $r_{end}-D\le  r \le r_{end} $
where $r_{end}$ is the $r$ value of the outermost grid point and 
$D$ is an adjustable parameter representing the width of the sponge. 
$D$ is typically chosen to be a few times the wavelength
of the scalar radiation moving out. 

The code is tested with equilibrium configuration data 
(the zero perturbation case).  We confirm that the evolutions leave
the metric and BD fields perfectly static for several thousand
time steps. 
We also checked the code with taking a large
number of BD parameter $\omega_{BD}$ which would converge 
to the results in GR.

\section{Equilibrium sequences}\label{sec:equili}
Before starting the dynamical study of boson stars, we
describe briefly the equilibrium solutions in BD theory. 

The existence of ground state boson stars in BD theory
are reported in \cite{GJ} for  
the case of  $\omega_{BD}=6$. 
The masses of the boson stars become smaller than the corresponding
configurations in GR. They also show  the
effects of the interaction term, $\Lambda$. 
In \cite{CS97}, the existence of higher node (excited) states 
are reported. They find that if the observational limit on 
$\omega_{BD}$ is admitted then the obtained configurations 
are very similar to GR.
They also describe the stability of the ground 
states configurations using catastrophe theory. However, the whole 
discussion on the stability including the excited states is unconfirmed
upto now.  Also the dynamical behavior of this system are totally 
unknown.

\begin{figure}[h]
\setlength{\unitlength}{1in}
 \begin{picture}(3.75,2.2)
\put(0.0,0.0){\epsfxsize=3.0in \epsfysize=2.12in \epsffile{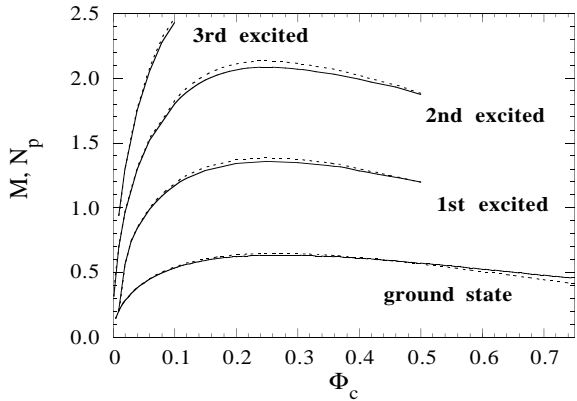} }
\end{picture}
\caption[fig-eq]{
Stable configurations of a boson star in Brans-Dicke theory. 
Masses of ground state and excited state stars upto 3 nodes 
are shown as functions of  the central matter field $\Phi_c$.
The solid lines and the dotted lines are the mass $M/(G_*/m)$ and the 
particle number $N_p/(G_*/m^2)$, respectively. 
}
\label{fig-eq}
\end{figure}

In Fig.\ref{fig-eq}, we show the sequences of equilibrium configuration of
boson stars in BD theory both for ground states and excited states 
upto three nodes.  These sequences are given by solving the set of 
equations described in \S \ref{sec2-equili}, and we applied 
the BD parameter $\omega_{BD}=600$ for plotting this figure. 
The total mass of boson star $M$ and its particle number $N_p$ 
(in unit $m$)
versus central boson density $\Phi_c$ profile is plotted. 
We see that, on the ground and excited state sequences, 
the signature of the 
binding energy $M-mN_p$ goes from negative to positive according to the
value of $\Phi_c$.
Configurations with positive binding energy are expected to be 
dispersive.

In the case of GR, we know 
the star ground state configurations 
to the left of the central maximum are {\it stable}. 
By `stable', we mean that under 
perturbations they move to new configurations on the same branch (left
of the maximum). 
We naturally suspect that the branch to the left of the maximum in the 
BD profile is also stable.
Therefore, in this and all further sections of this paper, 
we will refer to ground state boson
star configurations to the left of the maximum mass in Fig.\ref{fig-eq} as
the $S$-branch configurations and others as  $U$-branch
 configurations.
From the catastrophe discussion\cite{CS97}, we also suspect 
that the $U$-branch configurations are unstable. 

\begin{figure}[h]
\setlength{\unitlength}{1in}
\begin{picture}(3.75,2.2)
\put(0.0,0.0){\epsfxsize=3.0in \epsfysize=2.12in \epsffile{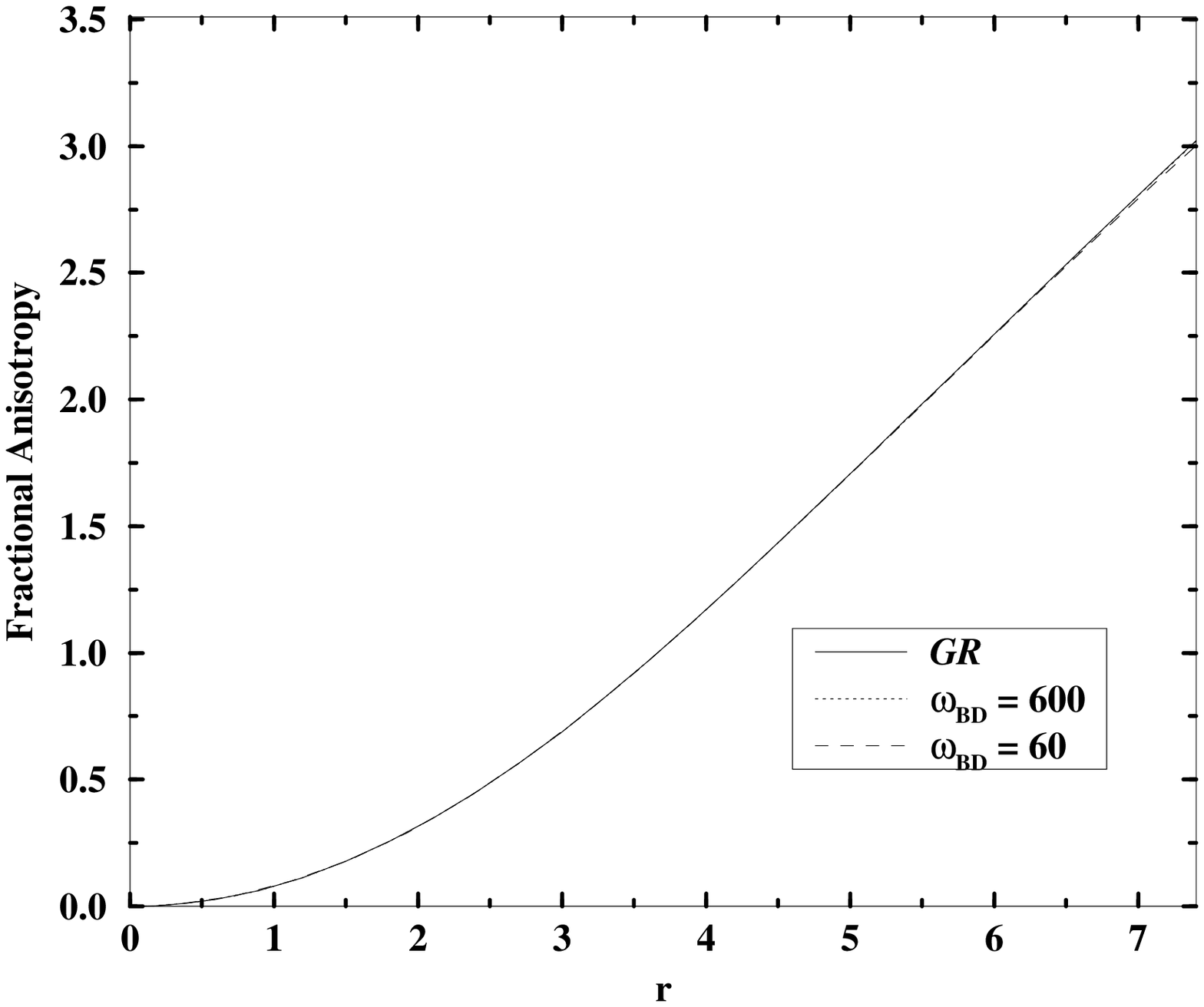} }
\put(0.5,0.7){\epsfxsize=1.1in \epsfysize=0.8in \epsffile{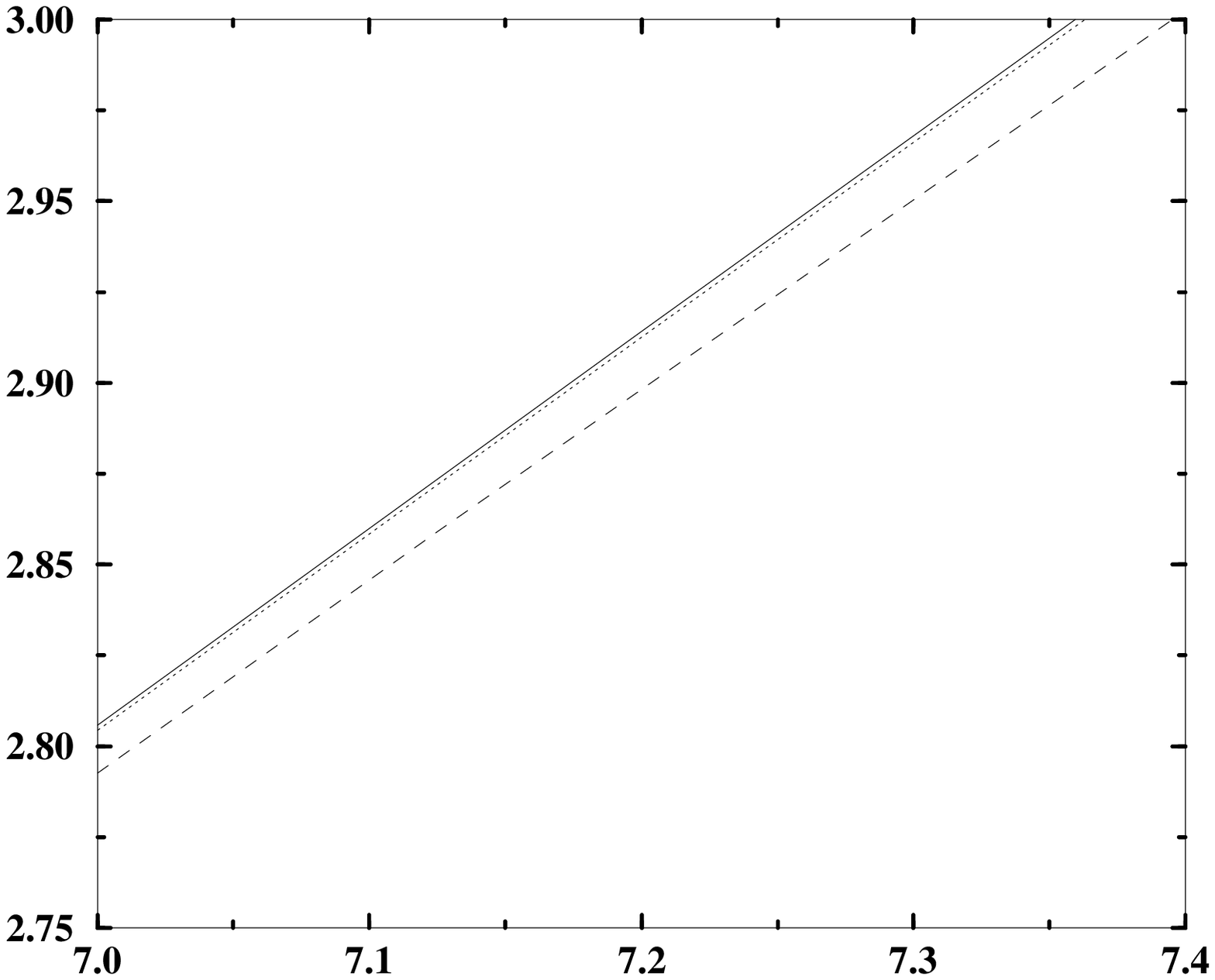} }
\end{picture}
\caption[fig-aniso1]{
Fractional anisotropy (\ref{fa}) is plotted for ground state 
configurations of central boson density $\Phi_c=0.2$ for GR and BD 
($\omega_{BD}=600$ and 60). The inset figure is the magnification of
the range $r=7.0$ to $7.4$.
}
\label{fig_aniso1}
\end{figure}
\begin{figure}[h]
\setlength{\unitlength}{1in}
\begin{picture}(3.75,2.2)
\put(0.0,0.0){\epsfxsize=3.0in \epsfysize=2.12in \epsffile{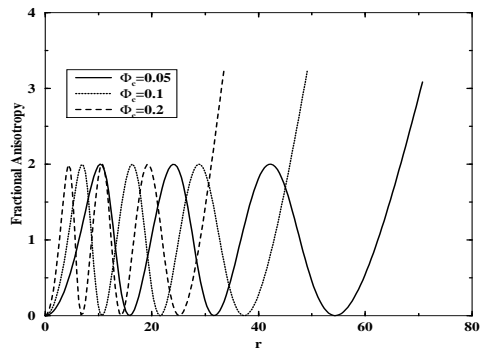} }
\end{picture}
\caption[fig-aniso2]{
Fractional anisotropy (\ref{fa}) is plotted for 3 node (third excited)
states in BD ($\omega_{BD}=600$) with different 
central boson densities $\Phi_c=0.05, 0.1$, and 0.2. 
The plots end at the radius of each star (95\% of the total mass). }
\label{fig_aniso2}
\end{figure}

The boson star system exhibits anisotropies due to the presence of
scalar fields.  Here, ``anisotropy" means the difference of 
the radial pressure from the tangential pressure in these 
configurations ($|T^r_r|\ne|T^\theta_\theta|$, where  $T_\mu^\nu$  is
the energy-momentum 
tensor defined as the right-hand-side of (\ref{einsteineq}).  
In the case of a scalar-tensor theory this
anisotropy could change from GR due to the presence of the additional
gravitational scalar field.
In the case of BD theory, we find the 
anisotropy slightly but not significantly lowers as compared to GR.
In Fig.\ref{fig_aniso1}, we show the fractional anisotropy 
\begin{equation}
f_a={(T^r_r-T^\theta_\theta)} / {T^r_r} \label{fa}
\end{equation}
for a ground state boson star with $\Phi_c=0.2$ for GR and BD with
$\omega_{BD}=600$ and 60. 
In Fig.\ref{fig_aniso2}, we show the fractional anisotropy 
for 3 nodes' (third excited) states of boson stars in BD
$\omega_{BD}=600$ with different $\Phi_c$.
At the nodes where the boson field derivative vanishes, all those 
fractional anisotropy are
close to 2, which is the analytically expected number.
This is because we can write the equilibrium
fractional anisotropy as
\begin{equation}
\frac{2\left[
 {({\partial_r}\varphi)^2
+e^{2a}  ( {\partial_r} \Phi ) ^2}\right]}
{
 ({\partial_r}\varphi)^2
+e^{2a} [ ( {\partial_r} \Phi )^2
+{g^2 \over N^2} \Phi^2 -2 g^2 e^{2a}
V(\Phi)]}.
\end{equation}
At the nodes $\Phi=0$ and as a result $V(\Phi)$ also vanishes.

The plots end at the radius of the boson
star
(we define the radius to be the $95\%$ mass radius of these
systems).
Consistent with Gleiser's results in the GR case \cite{gleiser}, we have
seen that the fractional
anisotropy at the radius of the stars are almost the same number for all
configurations. 
One of the consequences of
these anisotropies is that we cannot apply the adiabatic 
perturbation method to discuss
stability of this system, as discussed by Kaup\cite{K}.

\section{Stability of ground state equilibrium states}
\label{sec:stability}

In order to confirm whether the stars on the $S$-branch of the ground state 
in Fig.\ref{fig-eq} are stable, we show our simulation of
the perturbed configurations
of the $S$-branch stars in this section. 

Our perturbations
themselves simulate the annihilation (creation) of particles by enhancing
(reducing) the boson field smoothly in some region of the star. 
A typical perturbation is put in for
a region of the star from $r_1$ to $r_2$  and is of the form
\begin{equation}
\Phi_{perturbed} (r)=  \Phi_{original}(r) 
\left\{1  + p  \sin^n (\pi { r_2-r
\over r_2-r_1})\right\}
\end{equation}
where $p$ is the perturbation size and is positive for particle 
creation and negative for annihilation, and 
$n$ is an integer that can be varied. 

This perturbation
is put in at the initial time step and the lapse equation and
Hamiltonian equation are reintegrated to give new metrics for this scalar
field and original BD field. The momentum associated with the  BD
field is set to be $zero$ on the initial time slice. 

\subsection{Small perturbation of bosonic scalar field}
\begin{figure}[h]
\setlength{\unitlength}{1in}
\begin{picture}(3.75,2.0)
\put(0.375,1.0){\epsfxsize=2.5in \epsfysize=1.0in 
              \epsffile{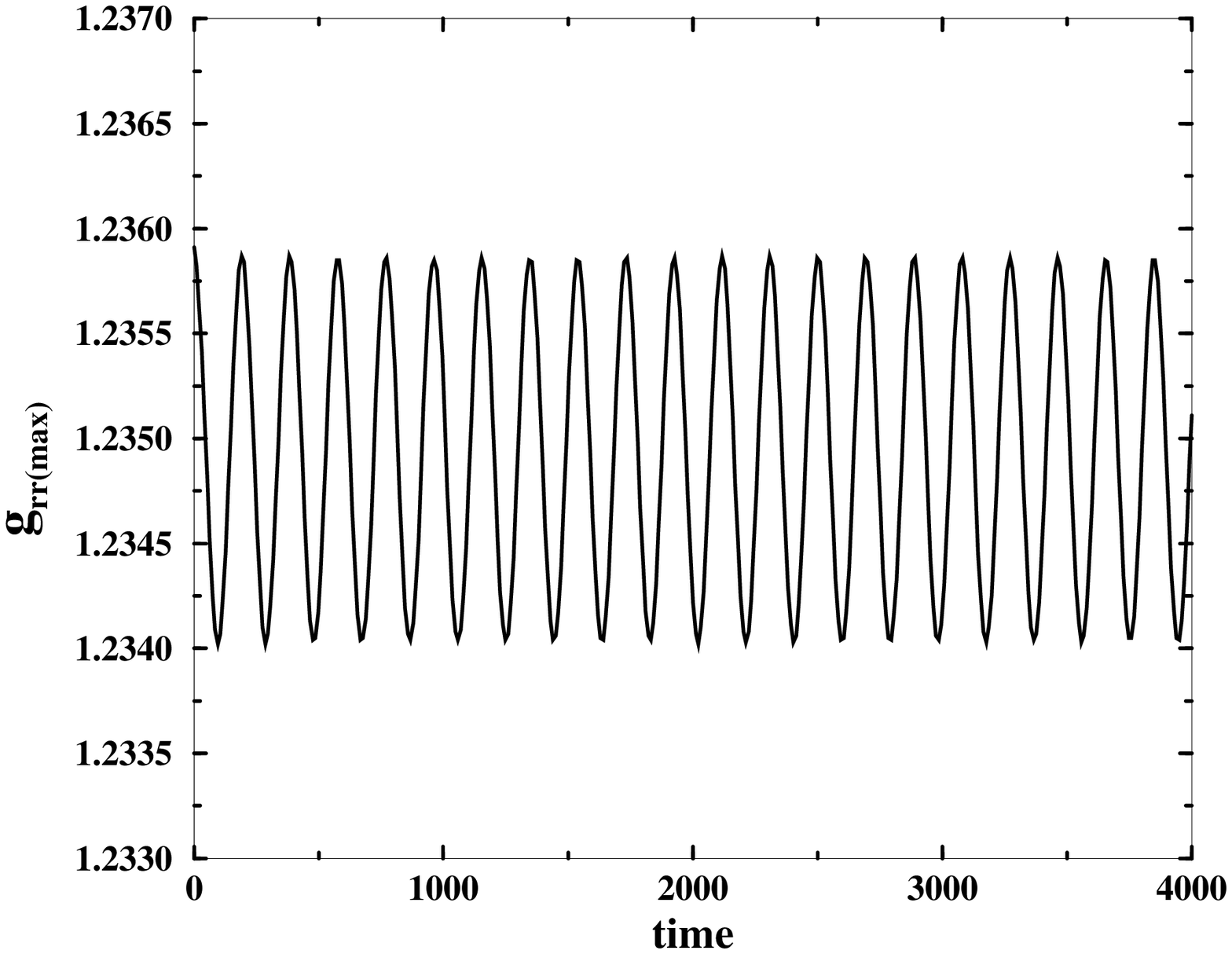} }
\put(0.375,0.0){\epsfxsize=2.5in \epsfysize=1.0in 
              \epsffile{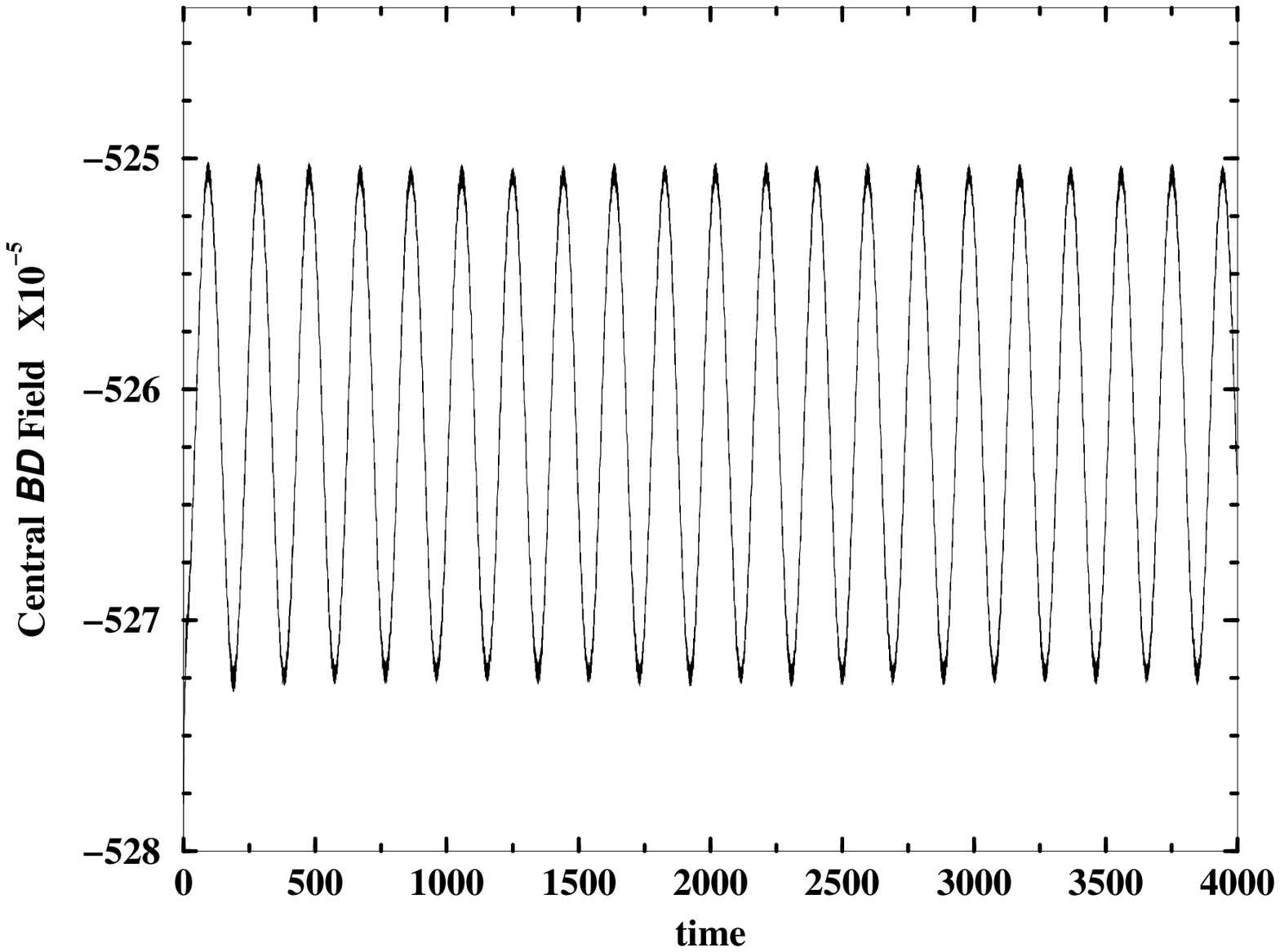} }
\end{picture}
\caption[stable_vphi]{
Quasinormal mode oscillation of a stable boson star.
The maximum value of the metric $g_{rr}$, 
and the central
Brans-Dicke field $\varphi(r=0)$
are plotted as functions of time. Both of them take on the  QNM
frequency of the star. The oscillation is 
virtually undamped for a long period of time.
}
\label{stable_grr}
\end{figure}

After we confirmed that our code retains its an equilibrium profile
for long time evolutions, we start our evolutions of an 
$S$-branch equilibrium state with a tiny perturbation. 
Here, `tiny' means that the difference of the mass between the original
and new configuration is less than $0.1\%$. 
We find that the system begins oscillating with a specific fundamental
frequency,
called a quasi-normal mode (QNM) frequency. 
Fig.\ref{stable_grr} shows the oscillations of 
the metric and the BD field 
for a boson star of central boson field density $\Phi_c=0.2$ and 
with $\omega_{BD}=600$. 
We see the BD field, as well as the metric
function, start oscillating at the same frequency.  
A Fourier transform of the metric and BD field data shows the overall 
frequency of the star to be $\frac{2}{\pi}\,0.03$ in the
nondimensional units. This is quite
close to the case in GR.

In GR, 
a QNM frequency increases as $\Phi_c$ become larger
 (radius become smaller) up to a point before starting to decrease 
rapidly 
to zero as it approaches the density corresponding to maximum 
mass signalling the onset of instability \cite{SS90}. 
We found the same feature in BD. 

\begin{figure}[h]
\setlength{\unitlength}{1in}
\begin{picture}(3.75,1.25)
\put(0.375,0.0){\epsfxsize=2.5in \epsfysize=1.0in  
\epsffile{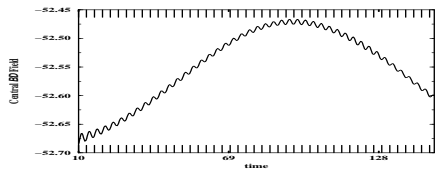} }
\end{picture}
\caption[stable_vphi]{
The magnification of the second figure of Fig.\ref{stable_grr}. 
 The time interval between
the ticks on the horizontal axis is $\pi$. 
We see the Brans-Dicke field has the underlying 
$\pi$ oscillation that it takes on. 
}
\label{stable_pi}
\end{figure}

In addition, the system takes on the proper underlying frequency,
which originates from the time dependence of the boson field $\Psi
\sim e^{i t}$.  
This frequency 
corresponds to $2\pi$ period in $t$ in our units, 
and appear in the metric and BD field 
 oscillation with $\pi$ period, 
from the structure of the equations. 
To show this feature we 
have enhanced the BD field oscillation (Fig.\ref{stable_grr})
in Fig.\ref{stable_pi}. The time interval between
the ticks on the horizontal axis is $\pi$.

\subsection{Large perturbation of bosonic scalar field}

Under large perturbations,  a stable boson star in GR expands
and contracts,  losing its mass at each expansion.  The
oscillations damp out in time
and the system finally settles down
into a new configuration on the $S$-branch. These features 
are now observed in BD theory. 

We show the effects of a large perturbation on a stable
BD boson star described above. 
The example we present is the case of initial data
with $\Phi_c=0.2$, mass $M=0.540G_*/m$ after perturbation 
(about $13\%$ lower in mass compared to unperturbed equilibrium 
configuration of $M=0.622G_*/m$).

\begin{figure}[h]
\setlength{\unitlength}{1in}
\begin{picture}(3.75,2.2)
\put(0.0,0.0){\epsfxsize=3.0in \epsfysize=2.12in 
              \epsffile{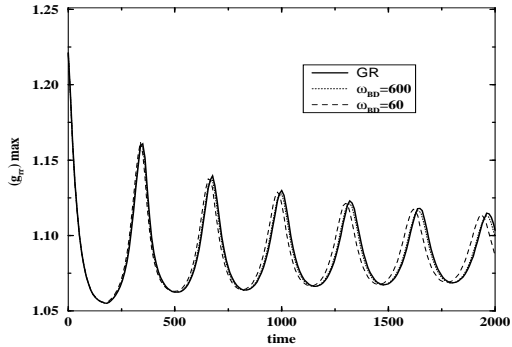} }
\end{picture}
\caption[stBS_grr]{
Finite perturbation of an $S$-branch boson star. 
Maximum metric $g_{rr}$ is plotted.
The metric is damped in time as the star settles to a new 
configuration.
}
\label{stBS_grr}
\end{figure}

\begin{figure}[h]
\setlength{\unitlength}{1in}
\begin{picture}(3.75,2.2)
\put(0.0,0.0){\epsfxsize=3.0in \epsfysize=2.12in 
              \epsffile{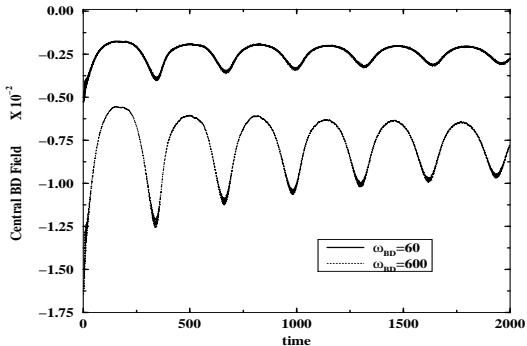} }
\end{picture}
\caption[stBS_vphi]{
The same model as Fig.\ref{stBS_grr}. 
 Central Brans-Dicke field $\varphi(r=0)$ is plotted. 
We see the oscillations damp out as the star settles, 
indicating again
transition to a new stable boson star configuration.
}
\label{stBS_vphi}
\end{figure}

The maximum radial metric $g_{rr}$ and the central BD field 
as a function of time are shown in Fig.\ref{stBS_grr} and 
Fig.\ref{stBS_vphi}, respectively. In both figures, we plotted
the case of $\omega_{BD}=600, 60$ as well as GR. 
The increase of the  maximum  $g_{rr}$ indicates 
the star is contracting,  reaching
its maximum value at the end of the contraction in a cycle. 
Then as the star expands,  the maximum  $g_{rr}$ decreases, 
reaching its
minimum at the end of the expansion.  
These processes repeat themselves
with the oscllations damping out in time as the star settles 
to a new stable
configuration with maximum radial metric of smaller
value than it started with (lower mass). The lower value
of the BD parameter shows a phase shift in comparison to GR 
which might be suggestive of a different rate of approach 
to the final configuration.
We see the same dynamical behavior in BD field (Fig.\ref{stBS_vphi}). 
It has the same oscillation frequency as that of
the metric. The oscillations damp out in time and the BD field settles
to a value closer to zero than it started at (lower final mass).

The system loses mass through 
radiation during its evolution. 
A comparison
of the mass as a function of time for BD case ($\omega_{BD}=
600$ and $60$ as well as GR shows little
 difference,  indicating
that the radiation is mostly scalar field radiation
and not scalar gravitational radiation. 
This is despite the BD field
oscillating in the BD case and being zero in the GR case.

\begin{figure}[h]
\setlength{\unitlength}{1in}
\begin{picture}(3.75,2.2)
\put(0.0,0.0){\epsfxsize=3.0in \epsfysize=2.12in
              \epsffile{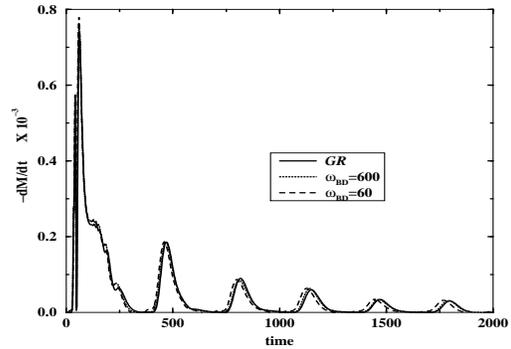} }
\end{picture}
\caption[stBS_dmdt]{
The same model as the previous figures. 
`Luminosity' $L=-dM/dt$  is plotted versus time. 
Clearly the radiation is decreasing in time.
}
\label{stBS_dmdt}
\end{figure}

The amount of mass radiated progressively decreases as 
can be seen in the luminosity profile
($-{dM}/{dt}$ versus time) shown in
Fig.\ref{stBS_dmdt}. 
Here again a comparison between GR and the
two BD parameters is shown. Again we see the phase shift for the
lower BD parameter (further from GR).
The system finally settles to a new state of smaller mass.

We have also conducted tests with perturbations that enhance 
the mass. For stable branch stars as
in the case of GR\cite{SS90} the star loses mass and settles 
to a new configuration on the
same branch.

\section{Evolution of $U$-branch stars and excited stars} 
\label{sec:transition}
Boson stars on the $U$-branch and excited states 
are inherently unstable in GR \cite{BSS}. 
Under perturbations that reduce the mass, boson stars on the
$U$-branch can migrate to the stable branch.  
In this section, we will see these features in BD theory too. 
They can be perturbed in a way so as to decrease their mass enough
that they migrate to new configurations on the $S$-branch.  
On the other hand, in GR, 
if they do not lose mass and migrate, then 
the boson stars of configurations
with $M<N_p\,m$
collapse to black holes.
Stars with $M> N_p\,m$ are dispersive and radiate out to infinity.
We will see also these features in BD theory.

\subsection{Migration to Stable Configuration} 
\begin{figure}[h]
\setlength{\unitlength}{1in}
\begin{picture}(3.75,2.2)
\put(0.0,0.0){\epsfxsize=3.0in \epsfysize=2.12in
              \epsffile{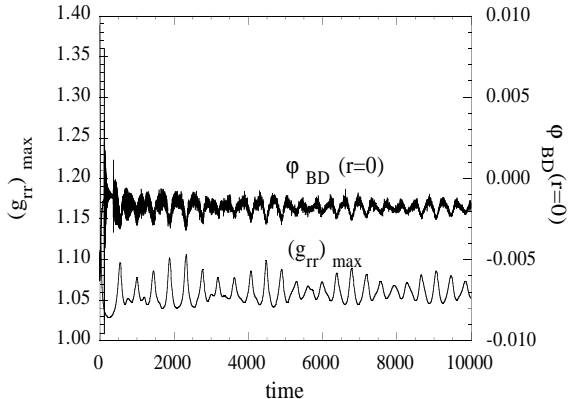} }
\end{picture}
\caption[trBSgrrvphi]{
Migration of an unstable boson star to a stable configuration; 
central Brans-Dicke field $\varphi(r=0)$ and maximum of 
$g_{rr}$ is plotted. 
There is a sharp initial drop in the radial metric as the 
star moves to the stable branch.
The oscillations damp out in time as the star settles.
}
\label{trBSgrrvphi}
\end{figure}

Our first dynamical example from $U$-branch boson stars is 
a migration process. 
As in GR, we have also seen migrations of these stars to the stable
branch when we remove enough scalar field smoothly from some
region of the star so as to decrease the mass by about $10\%$.
In particular, we show the migration of a star of central boson field
 $\Phi_c=0.35$ with unperturbed mass 0.625 $G_*/ m$. 
After perturbation, its mass is reduced to 0.558 $G_*/ m$. 
In Fig.\ref{trBSgrrvphi}, we show the maximum value of metric $g_{rr}$
and central value of BD field $\varphi(r=0)$ versus time.
The initial sharp drops in both  lines occur as the star
rapidly expands and moves to the stable branch.  
After that it oscillates and
finds a new configuration to settle into.  The damping of oscillations
as it settles is clearly seen in the figure.  
We see also  the BD field oscillations
damping out as the star gets closer and closer
to its final state.

\begin{figure}[h]
\setlength{\unitlength}{1in}
 \begin{picture}(3.75,2.2)
 \put(0.0,0.0){\epsfxsize=3.0in \epsfysize=2.12in
              \epsffile{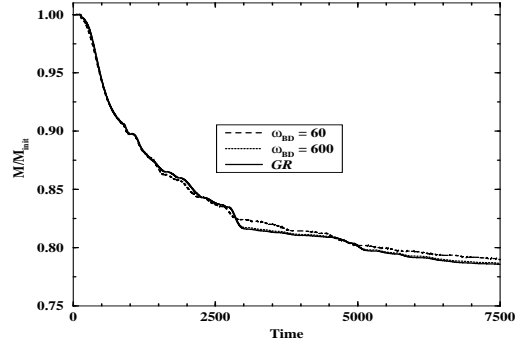} }
\end{picture}
\caption[trBS_mloss]{
Comparisons of total mass of the system $M$ rescaled by its initial
mass $M_{init}$
 during a migration process from $U$-branch 
star.  Three lines are plotted. Although the mass-loss is 
similar in the three cases the higher Brans-Dicke case 
clearly is closer to GR as it should be.
}
\label{trBS_mloss}
\end{figure}

The ratio of mass at time $t$ to the
initial mass for BD with parameter $\omega_{BD}=60$, $600$
and the GR case is shown in Fig.\ref{trBS_mloss}.  The flattening
of the curve at later time is indicative of the star settling down
to a new configuration.
Although convergence towards GR with increasing 
$\omega_{BD}$ is clearly
indicated,  there is no significant difference between the
three cases.  The amount of the total mass extraction from the system
is slightly supressed if we evolve in the BD theory.
By the time of 7500 shown in the plot, we see that the mass of the
star is
about 0.045 $G_{\ast}/m$, which corresponds to an equilibrium 
configuration with
$\Phi_c=0.06$, 
while our central density $\Phi_c$ is about  0.061,
 meaning that the star is quite
close to its final configuration.

\subsection{Transition to black hole} 
\begin{figure}[h]
\setlength{\unitlength}{1in}
 \begin{picture}(3.75,2.2)
 \put(0.0,0.0){\epsfxsize=3.0in \epsfysize=2.12in
              \epsffile{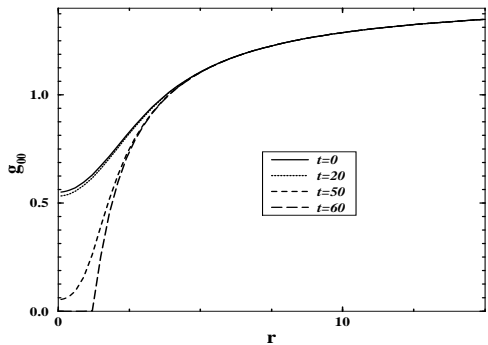} }
\end{picture}
\caption[shinka01]{
Dynamical transition from $U$-branch star to a black hole. 
The metric $g_{00}$ is plotted. The collapse of the lapse function 
is indicative of imminent black hole formation.
}
\label{trBH_g00}
\end{figure}

Contrary to the previous example, if we add a small mass to
$U$-branch stars, we can see the formation of a black hole in its 
evolution.
In Fig.\ref{trBH_g00}, we plotted an example of such evolution,
which indicates formation of a black hole. 
The initial data of this plot is boson star of central boson field 
$\Phi_c=0.35$, which is the same value with the previous migration case with 
an unperturbed mass of $0.625 G_{\ast}/m$).  
The system is perturbed very slightly so that the 
perturbed initial mass is $0.628 G_{\ast}/m$, which is 
less than $0.5\%$ greater than its initial mass. 
This is less than the maximum mass of a boson star of $0.632 G_{\ast}/m$
in the sequence of $\omega_{BD}=600$. 
The sudden collapse of the lapse function is
indicative of the imminent formation of an apparent horizon
due to the polar-slicing condition in our code \cite{bard}.
In addition to this the radial metric starts to blow up and the
code is no longer capable of dealing with the sharp gradients and
crashes. As an indicator of the
suddenness of the process, we see that in the configuration
shown the lapse has fallen to a value
of about $0.003$ by a time of 60 after being at $0.230$ at a time of $55$
and $0.5$ at a time of $50$ (the latter two points are not shown in the
plot).  
The lapse value at the edge of the grid is greater than one because
it has been scaled by the underlying frequency of
the system $\Omega$.

There is almost no loss in mass in this system and the time of 
collapse is quite similar to GR. 
In the GR case, we confirmed the
formation of black hole \cite{BSS_3D} very shortly after this point
($\approx 3\,M$ where $M$ is the mass of the system) by switching this data
into 3-dimensional code and evolving the system. 
Therefore we expect almost the same behavior in the BD case. 

Note that we have seen this behavior for denser boson stars with 
lower masses and the same
order of perturbation. This is true as long as they have $M<N\,m$.

The black hole formation in the BD theory is investigated by
Scheel, Shapiro and Teukolsky for dust collapse case \cite{SST}.
They found that the dynamical behavior of horizons (both event
horizon and apparent horizon) are quite different in the physical
Brans-Dicke frame, but the same as GR in the Einstein frame. 
Their results therefore also support our discussion of
the formation of black hole in the BD theory.

We can not show the emitted scalar waveform because we are using
polar slicing condition and can not proceed with the evolution after the
collapse of lapse. 
We are planning to evolve this system after the formation 
of black hole in 3 dimensional code.  The results will be reported
in the future. 

\subsection{Transition to ground state from excited state}

\begin{figure}[h]
\setlength{\unitlength}{1in}
 \begin{picture}(3.75,2.2)
 \put(0.0,0.0){\epsfxsize=3.0in \epsfysize=2.12in
              \epsffile{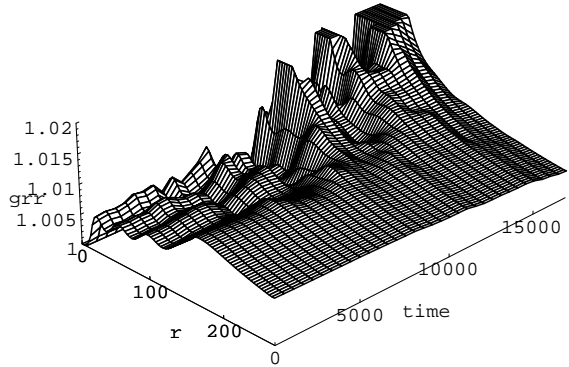} }
\end{picture}
\caption[shinka01]{
Dynamical transition from an excited state to a ground state 
 boson star configuration.  The metric  $g_{rr}$ is plotted.
The initial four peaks indicative of a 3-node star cascades to
the ground state after a long time evolution.
}
\label{node3grr}
\end{figure}

\begin{figure}[h]
\setlength{\unitlength}{1in}
\begin{picture}(3.75,2.2)
\put(0.0,0.0){\epsfxsize=3.0in \epsfysize=2.12in
              \epsffile{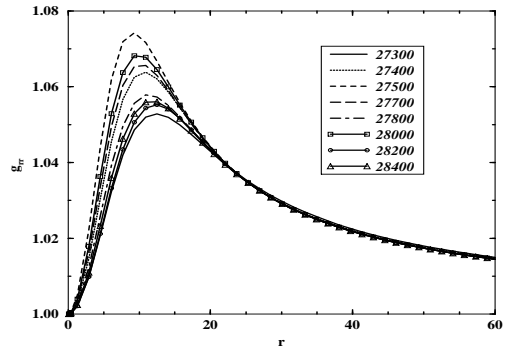} }
\end{picture}
\caption[node3grrend]{
Dynamical transition from an excited state to a ground state 
boson star configuration.  The metric  $g_{rr}$ is plotted
at later times to show its oscillations after the star reaches 
a ground state.
}
\label{node3grrend}
\end{figure}

Excited states of boson stars in general are not stable in GR case.
They form black holes if they
cannot lose enough mass to go to the ground state. We confirm the
same features for the BD case. 
In Fig.\ref{node3grr}, we plot the metric $g_{rr}$ of 
dynamical transition from an excited state with 3 nodes to a ground state 
boson star configuration.  
This star had an unperturbed mass of 3.249 $G_*/m$ corresponding
to a central density $\Phi_c=0.01$. 
After perturbation its mass was reduced to
0.919 $G_*/m$.  
The initial configuration has four metric maxima and the
final has one showing the transition. 
After it goes to the ground state, it oscillates and compactifies
to form a new configuration. 
We show the oscillations of the star
from a time of $27300$ to $28400$ in Fig \ref{node3grrend}. 
The $95\%$ mass radius
at this stage is about $100$ and the star has still to continue
its evolution for a while longer.
The amount of mass loss in this process is quite similar to the 
migration case (cf. Fig.\ref{trBS_mloss}).  
We also found that the difference of the theory is
little.   

Any configuration of an excited state star with a mass less than 
the maximum mass of a ground
state star is expected to go to the ground state. However since 
higher excited state $S$ branch
configurations have progressively greater masses for the same 
central
density ( Mass of $n$-nodes $> n-1  > \cdots >1 >$ ground
state boson stars) the configurations with masses less than 
the maximum ground state mass get
more and more dilute. 
We have seen the tendency of these
stars to migrate to the ground state in our tests. 
In \cite{BSS}, such cases have been described for
the GR case.
Since these dilute configurations correspond to large
stars with low oscillation frequencies, to evolve them until 
they go to the ground state is numerically
very costly as they take a very long time to do so. 
Configurations with slightly larger masses
may also be able to lose enough mass during their evolution 
and transit to the ground state. An exact
parameter search though would be very time consuming. 
We have rather chosen to just exemplify
the possibility of transitions with our example. We have 
taken a denser configuration to reduce the
time scale of evolutions and then perturbed it so much to 
ensure that no black hole forms. In fact we can
just think of the system as a perturbed distribution of 
bosonic fields that has 3 nodes and which
transits to the ground state rather than compare the 
perturbed system to the original unperturbed one.

\section{Formation of Boson Stars in Brans-Dicke theory}
\label{sec:formation}

In the previous sections, we analyzed boson stars in BD theory, starting
with those equilibrium or perturbed equilibrium configurations. 
However, we have not discussed whether or how such an equilibrium 
configuration actually forms in BD theory.
In this section, we answer this question by demonstrating the 
formation of boson star in BD theory. 
The formation of boson stars in GR has been discussed by 
Seidel and Suen\cite{SS94}.

We start our evolution with the initial data that has a 
Gaussian packet in the bosonic field $\Phi$, which represents a local 
accumulation of matter field:
\begin{equation}
\Phi= a \exp (-b x^2)
\end{equation}
where $a$ and $b$ are free parameters. 
We set the BD field $\varphi$ to be flat at the 
initial stage, so as to see if local inhomogeneity of the matter will
form a boson star in BD theory.  
We integrate 
the lapse equation (\ref{momco}) and Hamiltonian
constraint equation (\ref{branham}) to provide metric 
functons on the initial slice. 
We, then, evolve the set of dynamical equations 
(\ref{dyn-1})-(\ref{dyn-4}).

\begin{figure}[h]
\setlength{\unitlength}{1in}
\begin{picture}(3.75,2.2)
\put(0.0 , -0.125){\epsfxsize=3.0in \epsfysize=2.12in
\epsffile{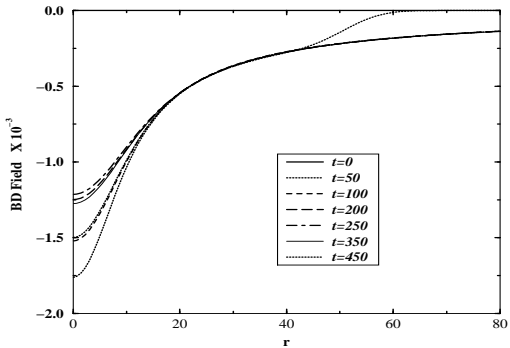}}
\end{picture}
\caption[fig_formC_2]{
An example of formation of boson star in Brans-Dicke theory. 
Snapshots of $\varphi_{BD} (r)$ are plotted for initial stage of
evolution.
}
\label{fig_formC_1}
\end{figure}

\begin{figure}[h]
\setlength{\unitlength}{1in}
\begin{picture}(3.75,2.2)
\put(0.1 , -1.9){\epsfxsize=3.3in 
\epsffile{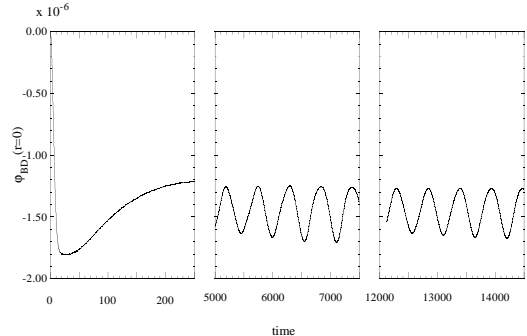}}
\end{picture}
\caption[fig_formC_2]{
An example of formation of boson star in Brans-Dicke theory. 
Dynamical behavior of the 
Brans-Dicke scalar field $\varphi_{BD} (x=0)$ is
 plotted for three evolution regions: 
earlier time, middle time and later time. We can see the field 
settling down to an equilibrium configuration (periodic oscillation).
}
\label{fig_formC_2}
\end{figure}

We find that, with particular parameters $a$ and $b$, this 
system actually forms a stable,  equilibrium configuration, which
might be recognized as the formation of a boson star. 
As a demonstration, we here show an evolution with parameters $a=0.1$
and $b=0.025$. The BD parameter $\omega_{BD}$ is taken to be $600$.
In Fig.\ref{fig_formC_1}, we show the BD field $\varphi$ as a function 
of radial coordinate at various earlier times of evolution. 
We see that the BD field becomes negative quickly 
and begins oscillating around a particular value.  
Actually the reader will find the BD field $\varphi$ is jumping around
$r=40-50$ at $t=50$.  The long time evolution of BD field is shown in
Fig.\ref{fig_formC_2}, in which we show the 
BD scalar field $\varphi_{BD}$ at the center 
for earlier times, middle times and later times. 
We can see the field 
settling down to a periodic oscillation in final phase, like in
the migration and transition cases in the previous section. 
The initial mass of this configuration in units of
$G_{\ast}$ is $0.39$ and the final(at the end of our simulation) 
about $0.384$.  At this stage the magnitude of the central
boson field is oscillating between 0.032 and 0.048. The BD field 
oscillates
between $-0.00126$ to $-0.00166$. This range of boson oscillations
corresponds to masses between 0.342 and 0.410 respectively while 
the BD field oscillations give
a mass between  .355 and .405 respectively. 
Given that the mass at this
stage is $0.384$ (consistent with the above) we expect that 
the final mass will be between $0.355$ and $0.384$.

\begin{figure}
\setlength{\unitlength}{1in}
\begin{picture}(3.75,2.2)
\put(0.1 , -1.9){\epsfxsize=3.3in 
\epsffile{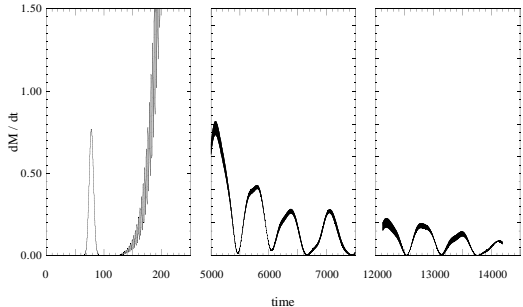} }
\end{picture}
\caption[fig_formC_3]{
Similar to Fig.\ref{fig_formC_2}, 
the emitted luminosity $L=-dM/dt$ is plotted. 
In the first evolution stage, the luminosity data takes on the 
underlying boson field square oscillation, after
having emitted one scalar pulse, related to the initial field 
configuration:
Gaussian pulse. The amount of mass loss decreases in time as 
the formed star settles.
}
\label{fig_formC_3}
\end{figure}


We show the luminosity $L (=-{dM}/{dt})$ versus time $t$ curve in 
Fig.\ref{fig_formC_3} within the same periods as in 
Fig.\ref{fig_formC_2}.
In the early stage, we see one pulse is emitted from the system. 
This is related to the outgoing pulse from our initial boson field
setting. 
After this initial pulse the system slowly takes on 
the characteristic $\pi$ oscillation
of $|\Psi|^2$ of the star as the star begins forming. 
After that,  
we see $L$ begins damped oscillations versus $t$. 
(We cut out the initial large amplitude luminosity around $t=200$). 
The system's evolution is followed for a long time, however, the
accuracy of the calculation is quite good with the hamiltonian 
constraint satisfied to
order $10^{-7}$ or better.

We also note that certain parameters $a$ and $b$ 
(mentioned at the beginning of this section)
will result in formation but others will not.
If we choose large amplitude $a$ and small $b$, then 
the initial configuration has too large a mass and is not 
dispersive enough
resulting in black hole formation during evolution. 
In the opposite limit if we have a very narrow localized wave
packet it has a tendency to be dispersive (wave equation). 
So if $b$ is too large
no boson star forms.
Intermediate between these two are configurations that form stable stars.
For example, if $a=0.1$,
the $b=0.01$ a black hole is formed, $b=0.025$ results in boson star 
formation as we
have shown, and  $b=0.035$ become flat space at the end of the evolution. 
On the other hand for $b=0.01$ and $a=0.05$ a boson star forms.


\section{Concluding Remarks} \label{sec:conclude}
We studied dynamical features of boson stars in the
Jordan-Brans-Dicke (BD) theory of gravity. 
By evolving the system numerically, 
we discussed the stability of those ground state $S$-branch equilibrium 
configuration, black hole formation and migration from $U$-branch 
solution, and 
transition process from excited state to ground state,
together with those scalar wave emissions (mass-loss from the
system). 

We showed that the basic features are the same as the general relativity
(GR) cases\cite{BSS}. 
Since we choose BD parameter $\omega_{BD}=600$ for most simulations
as we are interested in those observationable differences with GR, 
the signatures of boson stars such as oscillation frequency of the
final $S$-branch configurations and the emitted waves'  luminosity are
quite identical with GR. 
The scalar wave emissions are slightly supressed than GR. 
These results indicate to us that the scalar modes of 
gravity in the boson star system is already 
dominated by the boson scalar field, thus additional gravitational
scalar field does not change the dynamical behaviour
in BD theory of gravity. 

We also demonstrated the formation of a boson star from a Gaussian 
packet of bosonic field and flat BD field.  This success suggests to us
that the boson star is realizable object even in the BD
theory, opens windows to study them in astrophysical roles including
similar nontopological solitonic objects.

By showing that the $U$-branch of ground state equilibrium is unstable, 
we have confirmed 
the stability discussions by Comer and Shinkai for 
BD ground state using catastrophe theory. 
Our next problem is to check their prediction in the scalar-tensor
theory, especially Damour-Nordtvedt attractor model\cite{DN}, with which they
predict that the boson stars in the early universe will not be
formed. 
We are also now studying the existence of  ``oscillatons"\cite{SS91}, 
a self-gravitating solitonic object made from a real scalar field,  in the
scalar-tensor theory.   
These results will be reported elsewhere.

\acknowledgements
We thank Ed Seidel and Wai-Mo Suen for letting us modify their general 
relativity boson star code \cite{SS90} to scalar-tensor gravity.
We also thank Greg Comer and Clifford Will for useful conversations. 
We again appreciate Wai-Mo and Greg for useful comments on our drafts. 
This work was partially supported by the grant NSF PHYS 96-00507, 
96-00049, and NASA NCCS 5-153.


\vfill

\end{document}